\documentclass{aa} 
\usepackage{graphicx}
\usepackage{txfonts}
\usepackage{natbib}
\usepackage{color}
\usepackage{array}
\usepackage{subfig}

\newcommand{\fig}[1]{Fig.~\ref{fig:#1}}
\newcommand{\tab}[1]{Tab.~\ref{tab:#1}}
\newcommand{\sect}[1]{Sect.~\ref{sec:#1}}
\newcommand{\nofig}[1]{\ref{fig:#1}}

\begin{document}

\title{Deep imaging of the shell elliptical galaxy NGC\,3923 with MegaCam}

\author{M. B\'{i}lek\inst{1}\fnmsep\inst{2}
\and
J.-C. Cuillandre\inst{3}
\and
S. Gwyn\inst{4}
\and
I. Ebrov\'{a}\inst{1}
\and
K. Barto\v{s}kov\'{a}\inst{1}\fnmsep\inst{5}
\and
B. Jungwiert\inst{1}
\and
L. J{\' i}lkov{\' a}\inst{6}
}
\institute{Astronomical Institute,  Czech Academy of Sciences, Bo\v{c}n\'{i} II 1401/1a, CZ-141\,00 Prague, Czech Republic\\
\email{bilek@asu.cas.cz}
\and
Faculty of Mathematics and Physics, Charles University in Prague, Ke~Karlovu 3, CZ-121 16 Prague, Czech Republic
\and 
Laboratoire AIM Paris-Saclay, CEA/Irfu/SAp -- CNRS -- Universit{\' e} Paris Diderot, 91191 Gif-sur-Yvette Cedex, France, and Observatoire de Paris, PSL Research University, France
\and
Canadian Astronomy Data Centre, 5071 West Saanich Road, Victoria, BC, V9E 2E7, Canada
\and
Department of Theoretical Physics and Astrophysics, Faculty of Science, Masaryk University, Kotl\'a\v{r}sk\'a 2, CZ-611\,37 Brno, Czech Republic
\and
 Leiden Observatory, Leiden University, P.O.~Box~9513, NL-2300RA Leiden, The~Netherlands 
}

\date{Received ...; accepted ...}

\abstract
{The elliptical galaxy NGC\,3923 is known to be surrounded by a~number of stellar shells, probable remnants of an accreted galaxy. Despite its uniqueness, the deepest images of its outskirts come from the 1980s.   On the basis of  the modified Newtonian dynamics (MOND), it has recently been 
 predicted that a~new shell  lies in this region. }
{We obtain the deepest image ever of the {galaxy, map the tidal features in it, and  search for the predicted shell}.}
{The image of the galaxy was taken by the MegaCam camera at the Canada-France-Hawaii Telescope in the $g'$ band. It reached the surface-brightness limit of 29\,mag\,arcsec$^{-2}$. In addition, we reanalyzed an archival HST image of the galaxy.}
{We detected up to 42 shells in NGC\,3923. {This is by far the highest number among all shell galaxies.} We present the description of the shells and other tidal features in the galaxy. A~probable progenitor of some of these features was discovered. The shell system likely originates from two or more progenitors. The predicted shell was not detected, but {the new image revealed} that the prediction was based on incorrect assumptions and poor data. }
{}


\maketitle

\section{Introduction} \label{sec:intro}

Stellar shells in elliptical galaxies are likely remnants of galactic interactions (see \citealp{DC86,ebrovadiz}; or \citealp{ bilcjp} for a~review). The most accepted formation scenario is the accretion of a~small satellite along an almost radial trajectory \citep{quinn83} where the stars forming the shells are released from the progenitor when it passes through the pericenter. These stars form the shells when they reach their apocenters. For the Quinn model \citet{DC87} showed  that the progenitor has to make a~few oscillations before it dissolves to explain the high radial range (i.e., the ratio of the radii of the outermost and the innermost shell) observed in many shell galaxies. The accreted galaxy loses orbital energy by dynamical friction. The outermost shells are created first and the innermost shells later when the progenitor loses enough energy. Shells that were created together during a~particular pericentric passages are called a~generation (see \citealp{bil13} for details).

\citet{quinn84} pointed out the connection of the shell radii with the gravitational potential of the host galaxy. \citet{bil14a} (hereafter B14)  predicted the existence of an as yet unobserved shell in the elliptical galaxy NGC\,3923 at $\sim$2000\arcsec southwest (around 200\,kpc) from the galaxy's center. The prediction was made supposing the modified Newtonian dynamics (MOND) \citep{milg83a}, a~modification of Newtonian dynamics for low accelerations that eliminates the need for dark matter. This prediction was based on the shell radii {tabulated} by \citet{prieur88} and \citet{sikkema07}. If the shell were to be discovered, it would be the first discovery of an object predicted by MOND, a~test of MOND to exceptionally low gravitational acceleration for an elliptical galaxy, and the discovery of the biggest shell ever. 

Here we present the~new deep observation of the galaxy we obtained to search for the predicted shell. The observations of the outer parts of the galaxy (above a radius of 140\arcsec) supersede the best existing images so far coming from the 1980s. Moreover, we reanalyze the HST image obtained by \citet{sikkema06} which we use to look for the shells in the inner parts (under 140\arcsec) of the galaxy. 
We present a~new table of shell radii and describe the other tidal features in the galaxy{ that can serve to  constrain the gravitational potential (e.g., \citealp{bil13})}. The predicted shell is not detected, but we find that the analysis of B14 was based on incorrect assumptions and poor data. {In a forthcoming paper, we will present a quantitative photometric analysis.}

The distance of the galaxy from Earth is between 19.9 to 24\,Mpc according to the direct measurements in the NASA/IPAC Extragalactic Database. This corresponds to a linear scale of around 1\,kpc per 10\arcsec.

The paper is organized as follows. In \sect{obs} we describe observations, the data reduction, and the methods used
to enhance fine structures present in the galaxy. These structures are described in \sect{res}. Implications of the observation are discussed and concluded in \sect{disk}. { Appendix~\ref{sec:images} contains detailed images of all shells and other features found in  the galaxy. In Appendix~\ref{sec:minmask}, we describe our minimum masking method for the shell detection in detail.}

\section{Observations and data reduction} \label{sec:obs}

\begin{table}[htbp]
\caption{Mosaic tile centers}
\label{tab:mos} \centering
\begin{tabular}{ccc}\hline\hline
Tile & R.A. & Dec. \\\hline
1 & 11:49:30 & -29:11:00  \\
2 & 11:52:30 & -28:25:00 \\
3 & 11:49:30 & -28:25:00 \\
4 & 11:52:30 & -29:11:00\\\hline
\end{tabular}
\end{table}

\subsection{New observations by MegaCam} 
Our new observations of NGC\,3923 were carried out using the MegaCam imager at the 3.6\,m Canada-France Hawaii Telescope (CFHT) within the program 14BO12. They took place in January and February 2015. The galaxy was covered by a~four-tile mosaic (see \tab{mos}) with {the resulting field of view of around }$2\times2$\degr\ centered on the galaxy. 
Each tile is a~stack of 21 subexposures with the exposure time of 195\,s in the $g'$ band. The total exposure time was 273\,min. The pointing offsets between the subframes were 7\arcmin. These large offsets were required to subtract the parasitic light scattered within the instrument. It was necessary to fit the imaging of each tile to about 1\,hour to avoid substantial variation of the atmospheric illumination. \citet{atlasxxix} explain in detail the observation strategy needed to capture low-surface-brightness features with MegaCam. This camera has the angular resolution of 0.185\arcsec$/$pixel.

The raw data were processed by the Elixir pipeline at CFHT \citep{elixir} to deliver fully detrended images (standard CCD processing plus uniform zero-point illumination correction, and absolute  photometric and astrometric calibration). Those data were then fed into the low-surface-brightness branch of Elixir (Elixir-LSB, \citealp{elixirlsb, atlasxxix}) in order to subtract the large-scale component due to parasitic light scattered within the instrument. Elixir-LSB restores the true sky background to a~flatness level lower than 0.2\% of the absolute sky level (min-max), that is more than 7 magnitudes fainter, allowing the detection of extended features as faint as  29\,mag\,arcsec$^{-2}$ against a~flat background. This limit (standard for all Elixir-LSB data; see \citealp{ferrarese12}) is established through the visual identification of the faintest parts of extended features in the image such as shells, tidal streams, or galactic cirri. All frames were stacked by  MegaPipe \citep{megapipe}.

\subsection{Older observations by HST} 
{We studied our MegaCam image together with an image taken by the Hubble space telescope (HST) capturing the center of the galaxy. The HST observation was presented by \citet{sikkema06} and \citet{sikkema07}. We obtained the data from the Hubble Legacy Archive\footnote{{\url{http://hla.stsci.edu/}}}. We worked with a~stack of the images taken by the ACS$/$WFC camera in the filters F814W (978\,s) and F606W (1140\,s).  This made the exposure time of 35\,min in total. The angular resolution of this image was 0.05\arcsec/pixel.  }

\subsection{Shell detection methods}\label{sec:methods}
Shells are typically very faint,  low-contrast features (the exact values of the surface brightness of the shells will be presented in a~forthcoming paper).  Their visibility benefits from some image processing for contrast improvement.  We looked for shells via a~visual inspection of images processed in one of the following ways:

\emph{No processing} --  Most of the reported shells are visible in the original FITS images without further processing, although many of them with difficulty.

\emph{Model subtraction} -- We fit the image of the galaxy by a~smooth analytic profile in Galfit \citep{penggalfit}. {We use a  S{\' e}rsic profile  or the sum of two S{\' e}rsic profiles, both combined 
with sky.} The brightest stars, shells, tidal features, and galactic cirri are masked out {before the fitting}. The resulting image is the difference between the original image and the model. The method is used in Figs.~\nofig{s39} and \nofig{tidal}--\nofig{progen}. {We use a special fit for every image.}

\emph{Minimum masking} -- A~median filter is first applied to the original image to smooth the data on small angular scales. Next, an erosion (minimum) filter is applied. The result is the difference between the original image and the filtered image.  {Our erosion filter is circular  and the median filter is square. The side of the square is the same as the radius of the circle. As far as we know, minimum masking has never been described before (further details in Appendix~\ref{sec:minmask}).} This processing does not affect the position of shell edges. Shell edges transform into bright filaments on {the dark} background {(in the positive image)}. Figures~\nofig{s6}--\nofig{s25} and \nofig{s41}--\nofig{hook} have been processed by this method. {In the figures, the term ``size'' refers to the radius of the erosion filter.{ We tried several filter sizes. This helped us to further confirm the existence of the shell, see \fig{shdet}.} }

\emph{Unsharp masking} -- The original image is convolved with a~Gaussian. { The result at every pixel is twice the original intensity minus the intensity in the convolved image.} We use the standard deviations of the Gaussian of 44.4, 22.2, and 11.1\arcsec\,for our MegaCam image. This method proved to be less effective than  minimum masking and  model subtraction since unsharp masking  produces dark rings around bright stars that disturb the images and complicate the detection of the shells. {This method helped us only to confirm the existence of some shells.}

Before  including a~shell in our list (\tab{sh}), it had to be visible in images processed by two different methods at least. 

We detected more shells in the HST image than \citet{sikkema07} because they used only the model subtraction method. Some shells become more pronounced by the minimum masking method. On the other hand, while the model subtraction clearly  reveals, for example,  shell S40  (Figs.~\nofig{s39} and~\nofig{dust}), this shell is barely visible after the minimum masking (\fig{s41}).

\section{Results}\label{sec:res}
\begin{figure*}[t]
\resizebox{184mm}{!}{\includegraphics{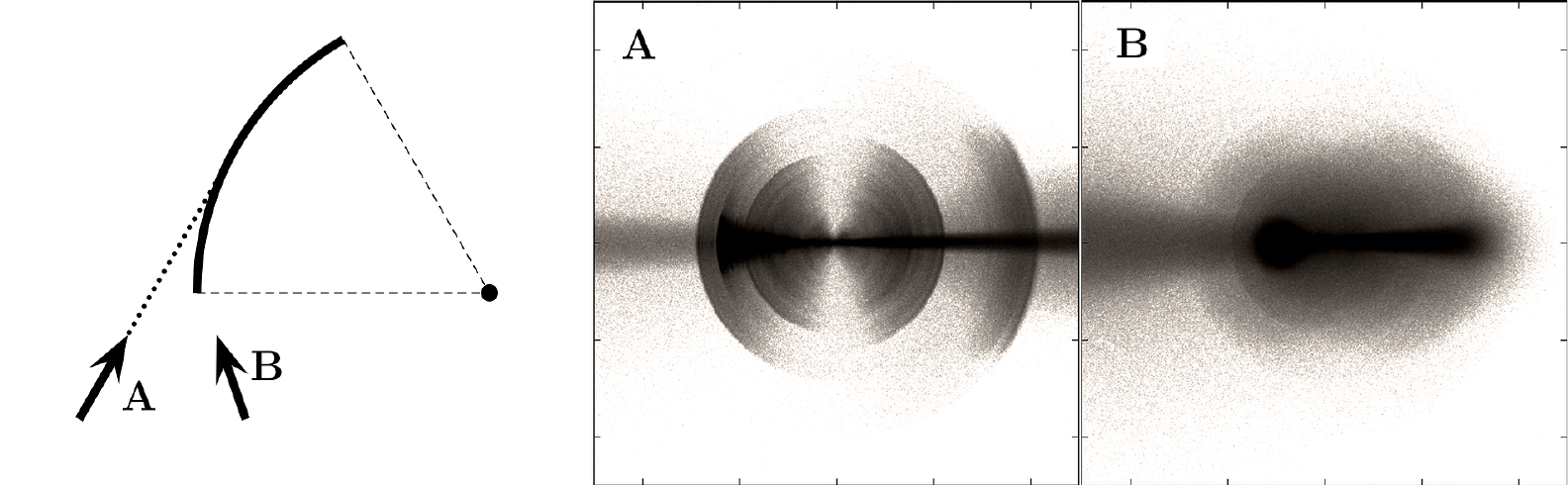}}
\caption{Shell appearance differs depending on the line of sight. In the
three-dimensional space, a~shell is a~part of a~spherical surface
centered on the galaxy. The observer in  direction $A$ sees the shell
as a~sharp-edged structure. The observer viewing the shell from 
direction $B$ does not see any part of the shell tangentially, and so 
the shell appears diffuse. The two images on the right show an illustrative simulation of a~radial minor merger between two spherical galaxies. Only the stars from the smaller galaxy are shown. The middle panel shows the view perpendicular to the line of collision. The view in the right panel is inclined by 35\degr\ to the line of collision.}
\label{fig:angle}
\end{figure*}

\subsection{Shells}
The shells or shell candidates we found are described in \tab{sh}, where we present the radius of the shell in arcseconds, its prominence degree,  an image processing method allowing its reliable detection, the {data} the shell was found in, the enumeration of the images showing shell, and notes. {The shells were detected by eye and thus the prominence rating is subjective. We state only the image processing method allowing a reliable detection with the minimal effort (order of difficulty: no processing, unsharp masking, minimum masking, model subtraction). Seeing a shell in an image processed by one method, we could note it in almost all the images processed in other ways.} The shell system is approximately axially symmetric. The symmetry axis coincides with the major photometric axis of the galaxy.  {Their position angle is close to 225\degr (the southwest-northeast direction).} 

The shells that have the radii with the plus signs in \tab{sh} lie on the northeastern half of the galaxy and those with minus signs on the southwestern half. We measured the radius of every shell along the major photometric axis of the galaxy.  We measured the radius from the brightest pixel in the center of the galaxy to the outer cut-off of the shell. { The assumed galaxy center had the FK5 coordinates of R.A.~=~11:51:01.7 and Dec.~=~-28:48:21.55.}

{In a forthcoming paper, the table will be completed by the shell equatorial coordinates, surface brightness estimates, and the shell azimuthal widths. This will be interesting for comparison with simulations.}

{In the center, we looked for the shells independently in the MegaCam and the HST image. The shells were more visible in the HST image because of its higher angular resolution and the absence of seeing. As can be noted from \tab{sh}, we missed several shells in the MegaCam image. All of them are seen as very diffuse in the HST image. On the contrary, we detected only one shell candidate in the MegaCam image that we missed in the HST image (radius of -87\arcsec). It would have been rated at the lowest prominence degree. In the HST image this feature looks like a short filament rather than a shell and so we do not include it into our table. For the shells captured in the HST image, the table states the radii measured from this image.}

All the shells {we detected} are shown in Figs.~\nofig{out}--\nofig{s41}. They are marked by~S$n$.

\subsection{Other tidal features} \label{sec:other}
Figures~\nofig{hook}--\nofig{progen} were processed to show the other tidal features in the galaxy. 

Shell S3 has a~much narrower azimuthal extent than the other shells in the galaxy. We can see that it lies at the end of a~stream extending approximately {from the core of NGC\,3923} (Figs.~\nofig{tidal}, \nofig{outsk}, and~\nofig{progen}). A~compact galaxy lies on the symmetry axis of this stream. The stream gets thinner close to the galaxy. It is likely the progenitor of the shell S3 and possibly of some others. The projected distance of this galaxy from the center of NGC\,3923 is 474\arcsec\ (around 47\,kpc supposing that it  lies close to NGC\,3923). It is  the elliptical galaxy PGC\,3097920. According to the HyperLeda\footnote{http://leda.univ-lyon1.fr/} database  \citep{hyperleda}, it has a heliocentric radial velocity of 1663$\pm$39\,km\,s$^{-1}$, while that of NGC\,3923 is 1550$\pm$15\,km\,s$^{-1}$, and so they can be  in  physical proximity.

A hook-like structure (H in  Figs.~\nofig{out}, \nofig{hook}, \nofig{tidal}, and \nofig{outsk}) lies northwest of the galaxy center. First, it goes to the north from the galaxy center, then it bends to the west and disappears pointing to the south. Similar structures can be produced by an accretion of a~disk galaxy as we can see in the simulations depicted in \citet{DC86}. On the other hand, this feature could also be a~galactic cirrus. The western end of the hook is located near a~triangular object (C1 in the figures) with a~filamentary structure typical for cirri. The hook itself appears smooth.

In the neighborhood of the hook and the shell S5, there is a~fan feature (F~in  Figs.~\nofig{out}, \nofig{tidal}, and \nofig{outsk}). It {is reminiscent of} a~shell, but its edge is not tangential to the direction toward the galaxy center and it has a~lower azimuthal extent than most of the shells in the galaxy.

Another prominent tidal feature is the stream (labeled ST in Figs.~\nofig{out}--\nofig{s9} and \nofig{hook}--\nofig{outsk}) south of the core of NGC\,3923. It points almost precisely in the north-south direction. It has no internal structure. It does not seem  to be an extension of the hook (\fig{hook}).  It disappears near the edge of the shell S4 on the south. At  first sight, it appears to be a~reflection caused by the star below it. {However, if it were an artifact, it would have to move with respect to the stars during our large offsets, but when we checked the individual subexposures, it was not the case.} Moreover, it is too fuzzy to be a~common artifact. 

As has been done in previous observations (e.g., \citealp{sikkema07,pence86}), we note several dust features near the center of the galaxy in our MegaCam image (\fig{dust}). {\citet{pence86} reported that the most prominent cloud (labeled  DP in \fig{s15}) contains H$\alpha$ emission with the same redshift as NGC\,3923. This proves that the dust lies in the galaxy and not in the foreground.}
 
\subsection{Galactic cirri}
We can see many  galactic cirri in our MegaCam image (labeled  C$n$). They are dust clouds in our own Galaxy. They  typically show  a fine filamentary structure. As is well known, galactic cirri can  easily be mistaken for tidal features in the external galaxies even if we have observations in several filters (e.g., \citealp{atlasxxix}). The angularly large cirri can be identified by comparing the optical images with the radiation maps in 857\,GHz obtained by the Planck spacecraft \citep{atlasxxix}. There is a~structure in the near-infrared map which could correspond to the stream. We did not identify such a~counterpart for any of the other tidal features we report, but the possibility of confusion is not fully excluded.  

\section{Discussion and conclusions}\label{sec:disk}
We presented our new deep image of the shell elliptical galaxy NGC\,3923. It was taken in the $g'$ band and reaches the surface-brightness limit of 29\,mag\,arcsec$^{-2}$. In addition, we reanalyzed an archival HST image of the galaxy. The motivation for our new observation was to look for the shell predicted by B14 to make a~test of the MOND theory of modified dynamics. The predicted shell was not found.

We detected 42 shells or shell candidates in total. Even though a~few candidates are highly doubtful, the number is  surprisingly higher than previously reported. The compilation by B14 contains only 27 shells. Two of the shells listed there allegedly lying at the radii $+630$\arcsec and $-520$\arcsec\ do not exist. {\citet{prieur88} possibly confused the halves of the galaxy. The shell at $+630$\arcsec\ is probably S4 with the radius of $-630$\arcsec\ and the shell at $-520$\arcsec\ is perhaps S5 with the radius of $+490$\arcsec. }This means that the prediction of B14 was based on incorrect and incomplete data. Next, this teaches us that many shells ($\sim$17) can easily escape observations, not a maximum of only six as assumed by B14.

A probable progenitor of at least one of the shells was found. It is the galaxy PGC\,3097920.  {Since the ratio of the projected radii of the innermost prominent shell S41 and this satellite is 26, the shell system was likely created by two or more progenitors. This follows from the findings from simulations summarized in \citet{DC87}. It
was found that the shell systems with a radial range higher than around 6 had to be made of multiple generations of shells (see \sect{intro}) in the framework of the minor merger model. The largest shells form first, while the inner shells form later when the secondary loses enough orbital energy by dynamical friction. But even if the observed progenitor is now in the apocenter, it has enough energy to create shells as large as some of the largest shells in NGC\,3923 in its next pericentric passage. Thus, it seems that the shells had to be created by more than one progenitor. This is supported by the fact that the number of shells in NGC\,3923 is the highest of all galaxies. The works of \citet{bil13} and B14 assumed that the whole system was created from one progenitor.}

For these reasons, our observation has no implication for the validity of MOND at this moment. The consistency of the shell distribution with MOND \citet{bil13} will have to be reanalyzed.

The shell system in NGC\,3923 held several records before our observation; it contained the largest known shell, the highest number of shells, and it had the highest radial range. After our observation, the galaxy becomes even more exceptional. We increased the number of shells and the radial range raised from 65 to 108. {There are several reasons for the substantial increase in the number of shells: 1) the depth of our MegaCam image, 2) other image processing methods applied to the HST image, and 3) more emphasis on the complete detection of shells motivated by the possibility of constraining the gravitational potential \citep{bil13}.}

We  note in the figures and \tab{sh} that the well-defined shells are mixed among the diffuse and very faint shells. There are several reasons for this. For example, the shells can come from different shell generations or from different progenitors. It is also possible that  the diffuse shells are shells seen from unfavorable angles. Indeed, shells are parts of spheres in three-dimensional space. If the shell is to be observed as a~sharp-edged feature, the line of sight must be tangential to the shell surface (\fig{angle}). In the opposite case, the shell edge appears diffuse or the shell is not observable.

\begin{acknowledgements}
We acknowledge support from the following sources:  project RVO:67985815 by the Czech Science Foundation (MB, IE, BJ, and KB);  grant MUNI/A/1110/2014 by  Masaryk University in Brno (KB); and  project SVV-260211 by Charles University in Prague (MB). The work of MB was supported by the European Commission through project BELISSIMA (BELgrade Initiative for Space Science, Instrumentation and Modelling in Astrophysics, call FP7-REGPOT-2010-5, contract no. 256772). MB thanks Pavel Kroupa for providing financial support through the University of Bonn.    
Based on observations obtained with MegaPrime/MegaCam, a~joint project of CFHT and CEA/DAPNIA at the Canada-France-Hawaii Telescope (CFHT), which is operated by the National Research Council (NRC) of Canada, the Institut National des Science de l'Univers of the Centre National de la Recherche Scientifique (CNRS) of France, and the University of Hawaii. 
The research leading to these results has received funding from the European Community's Seventh Framework Programme (FP7/2013-2016) under grant agreement number 312430 (OPTICON). 
We acknowledge the usage of the HyperLeda database (http://leda.univ-lyon1.fr). 
Based on observations made with the NASA/ESA Hubble Space Telescope, and obtained from the Hubble Legacy Archive, which is a~collaboration between the Space Telescope Science Institute (STScI/NASA), the Space Telescope European Coordinating Facility (ST-ECF/ESA) and the Canadian Astronomy Data Centre (CADC/NRC/CSA). 
\end{acknowledgements}

\bibliographystyle{aa}
\bibliography{citace}

\newcolumntype{P}[1]{>{\centering\arraybackslash}p{#1}}
\begin{table*}[htbp]
 \centering
 \caption{Detected shell candidates in NGC\,3923}
   \addtolength\tabcolsep{-1pt}
   \renewcommand\arraystretch{1.5}
\begin{tabular}{p{0.04\linewidth}P{0.09\linewidth}P{0.1\linewidth}P{0.08\linewidth}P{0.07\linewidth}lp{0.39\linewidth}}\hline\hline
Label & Major-axis radius [\arcsec] & Promi\-nence & Data & Detection method & Figures & Note \\\hline
S1 & +1170 & 2 & MC & 1 & \nofig{out} & largest \\
S2 & -952 & 4 & MC & 1 & \nofig{out}, \nofig{s6} & MP, 907-1012\arcsec, highly uncertain, possibly merged star halos \\
S3 & -846 & 1 & MC & 1 & \nofig{out}, \nofig{s6}, \nofig{hook} & narrow, connected with the ``progenitor'' \\
S4 & -630 & 3 & MC & 1 & \nofig{out}, \nofig{s6}, \nofig{hook} & MP, 640-590\arcsec, irregular, diffuse, near  the end of the ``stream'' \\
S5 & +490 & 3 & MC & 1 & \nofig{out}, \nofig{s9}, \nofig{hook} & MP, possibly other tidal feature, small azimuthal extent, opposite S8, connected with the ``hook'', shells at lower radii have surface brightness increased at the azimuth of this shell \\
S6 & -430 & 4 & MC & 3 & \nofig{s6}, \nofig{s9} & MP, very diffuse and faint, highly uncertain \\
S7 & +363 & 1 & MC & 1 & \nofig{out}, \nofig{s9}, \nofig{hook} &  \\
S8 & -338 & 3 & MC & 3 & \nofig{s6}, \nofig{s9}, \nofig{hook} & {MP,  not on axis, only in the eastern half, opposite  S5, shells at lower radii are brighter on the azimuth of S8} \\
S9 & +331 & 3 & MC & 3 & \nofig{s9} & {copies the edge of S7 but lower azimuthal extent} \\
S10 & -280 & 1 & MC & 1 & \nofig{s9}, \nofig{hook} &  \\
S11 & +260 & 3 & MC & 3 & \nofig{s9} & MP, very diffuse,  a~smooth step in surface brightness, possibly other tidal feature  \\
S12 & -237 & 4 & MC & 3 & \nofig{s9} & MP, sharp-edged, small azimuthal extent, possibly other tidal feature \\
S13 & +203 & 1 & MC & 1 & \nofig{s9}, \nofig{hook} &  \\
S14 & -150 & 1 & MC & 1 & \nofig{s9} &  \\
S15 & +146 & 2 & MC & 1 & \nofig{s9}, \nofig{s15} &  \\
S16 & +130 & 1 & MC, HST & 1 & \nofig{s15} &  \\
S17 & -120 & 3 & MC, HST & 1 & \nofig{s15} & MP \\
S18 & -103 & 1 & MC, HST & 1 & \nofig{s15} &  \\
S19 & +99.5 & 3 & MC, HST & 3 & \nofig{s15} &  \\
S20 & -92.0 & 4 & HST     & 3 & \nofig{s15} & MP, very faint and diffuse, smooth step in surface brightness, possibly other tidal feature \\
S21 & +90.6 & 3 & MC, HST & 3 & \nofig{s15} & MP \\
S22 & -78.4 & 2 & MC, HST & 1 & \nofig{s15}, \nofig{s25} & sharpness varies with azimuth \\
S23 & +72.8 & 2 & MC, HST & 1 & \nofig{s15}, \nofig{s25}, \nofig{s39} &  \\
S24 & -71.4 & 3 & MC, HST & 3 & \nofig{s15}, \nofig{s25} & MP, not on axis, only in the western half,  diffuse, possibly other tidal feature \\
S25 & -68.6 & 1 & MC, HST & 1 & \nofig{s15}, \nofig{s25} &  \\
S26 & +64.8 & 3 & MC, HST & 3 & \nofig{s15}, \nofig{s25} &  \\
S27 & +60.0 & 2 & MC, HST & 1 & \nofig{s15}, \nofig{s25} &  \\\hline
\end{tabular}\label{tab:sh}
\tablefoot{ \textbf{\textit{Label}} -- Shell designation. \textbf{\textit{Radius}} -- Shell radius in arcseconds. Shells lying on the north-eastern side of the galaxy have the positive sign; the shells on the south-western side of the galaxy have the negative sign. \textbf{\textit{Prominence}} --  1 = Prominent and sharp-edged, 2 = Prominent and diffuse-edged, 3 = Faint but probably existent, 4 = Questionable.   \textbf{\textit{Data}} -- The image where the shell was detected: our observation with MegaCam (MC) or the archival HST image \citep{sikkema06}. \textbf{\textit{Detection method}} -- A~method allowing a~good detection of the shell: 1 = No processing, 2 = Model subtraction, 3 = Minimum masking (\sect{methods}, {Appendix~\ref{sec:minmask}}). \textbf{\textit{Figures}} -- The figures showing the shell. \textbf{\textit{Notes}} -- MP = Missed by the previous observations.}
\end{table*}

\addtocounter{table}{-1}
\begin{table*}[htbp]
\caption{Continued}
   \addtolength\tabcolsep{-1pt}
   \renewcommand\arraystretch{1.5}
\begin{tabular}{p{0.04\linewidth}P{0.09\linewidth}P{0.1\linewidth}P{0.08\linewidth}P{0.07\linewidth}lp{0.39\linewidth}}\hline\hline
Label & Major-axis radius [\arcsec] & Promi\-nence & Data & Detection method & Figures & Note \\\hline\
S28 & -60.0 & 3 & MC, HST & 3 & \nofig{s15}, \nofig{s25} & MP, diffuse, possibly other tidal feature \\
S29 & -55.9 & 1 & MC, HST & 1 & \nofig{s25}, \nofig{s39} &  \\
S30 & +51.0 & 1 & MC, HST & 1 & \nofig{s25} & very prominent \\
S31 & -49.0 & 3 & MC, HST & 3 & \nofig{s25} & MP, diffuse, possibly other tidal feature \\
S32 & +45.4 & 3 & HST     & 3 & \nofig{s25} & MP, diffuse, probably the shell at 41.5\arcsec in \citet{sikkema07} \\
S33 & -44.7 & 1 & MC, HST & 1 & \nofig{s25}, \nofig{s39}, \nofig{s41} & very prominent \\
S34 & +38.0 & 3 & MC, HST & 3 & \nofig{s25} & MP, very faint on axis, well defined on the western half \\
S35 & -37.5 & 3 & MC, HST & 3 & \nofig{s25}, \nofig{s41} & diffuse \\
S36 & +34.8 & 1 & MC, HST & 1 & \nofig{s25}, \nofig{s39}, \nofig{s41} &  \\
S37 & -30.8 & 3 &     HST & 2 & \nofig{s39}, \nofig{s41} &  bright smooth step in surface brightness, possibly other tidal feature, probably the shell at -28.7\arcsec in \citet{sikkema07} \\
S38 & +29.5 & 1 & MC, HST & 1 & \nofig{s25}, \nofig{s39}, \nofig{s41} &  \\
S39 & -24.2 & 4 & MC, HST & 3 & \nofig{s39}, \nofig{s41} & MP, highly questionable, faint, small azimuthal extent, possibly edge of a~faint dust cloud \\
S40 & +19.8 & 1 & MC, HST & 2 & \nofig{s39}, \nofig{s41} & very prominent after model subraction, abrupt cutoff in azimuth \\
S41 & -18.1 & 1 & MC, HST & 1 & \nofig{s39}, \nofig{s41} & abrupt cutoff in azimuth \\
S42 & -10.8 & 3 & MC, HST & 3 & \nofig{s41} & MP, very faint  \\\hline
\end{tabular}
\tablefoot{ \textbf{\textit{Label}} -- Shell designation. \textbf{\textit{Radius}} -- Shell radius in arcseconds. Shells lying on the northeastern side of the galaxy have a positive sign; the shells on the southwestern side of the galaxy have a negative sign. \textbf{\textit{Prominence}} --  1 = Prominent and sharp-edged, 2 = Prominent and diffuse-edged, 3 = Faint but probably existent, 4 = Questionable.   \textbf{\textit{Data}} -- The image where the shell was detected: our observation with MegaCam (MC) or the archival HST image \citep{sikkema06}. \textbf{\textit{Detection method}} -- A~method allowing a~good detection of the shell: 1 = No processing, 2 = Model subtraction, 3 = Minimum masking (\sect{methods}, {Appendix~\ref{sec:minmask}}). \textbf{\textit{Figures}} -- The figures showing the shell. \textbf{\textit{Notes}} -- MP = Missed by the previous observations.}
\end{table*}

\newpage
\begin{appendix}
\section{Images}
\label{sec:images}
Here we present images of the individual features we found in our MegaCam image and in the archival HST image \citep{sikkema06}. In all images, north is to the top and east is to the left. The abbreviations used in the images are as follows: 
\begin{description}
        \item[S$n$] -- Shells.
        \item[A] --  Artifact caused by reflection of a~bright star. It is present only on some of the subexposures.
        \item[ST] -- Stream (see \sect{other}). It is present on all subexposures.
        \item[F] -- Fan (\sect{other}).
        \item[H] -- Hook (\sect{other}).
        \item[DP] -- Dust patch.
        \item[P] -- Progenitor of  shell S3 and possibly of some others.
        \item[C$n$] -- Probable galactic cirri. They show filamentary structure.
\end{description}

\begin{figure}
\includegraphics[width=\hsize]{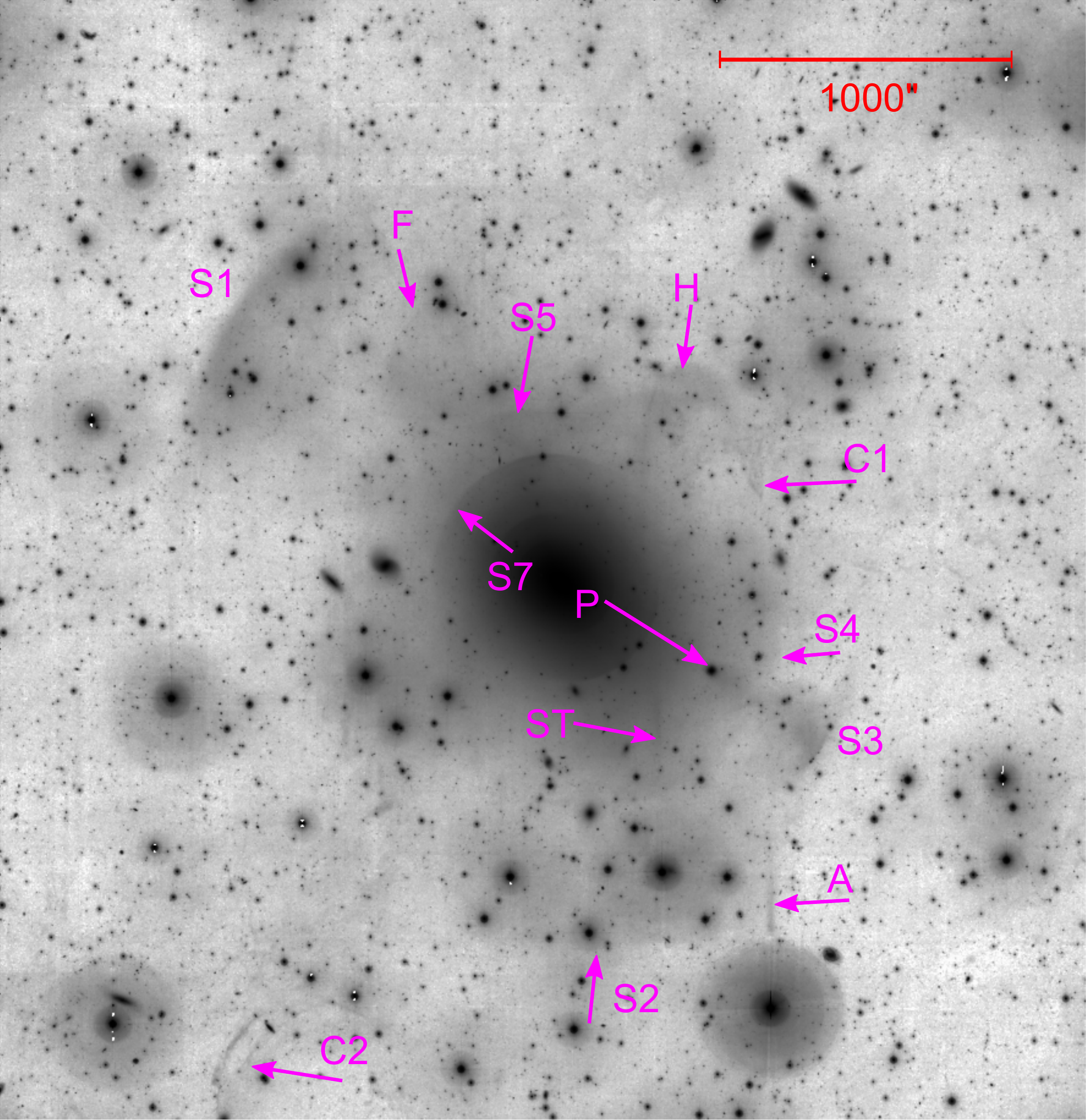}
\caption{MegaCam image. No processing.}
\label{fig:out}
\end{figure}

\begin{figure}[b!]\vspace{6ex}
\includegraphics[width=\hsize]{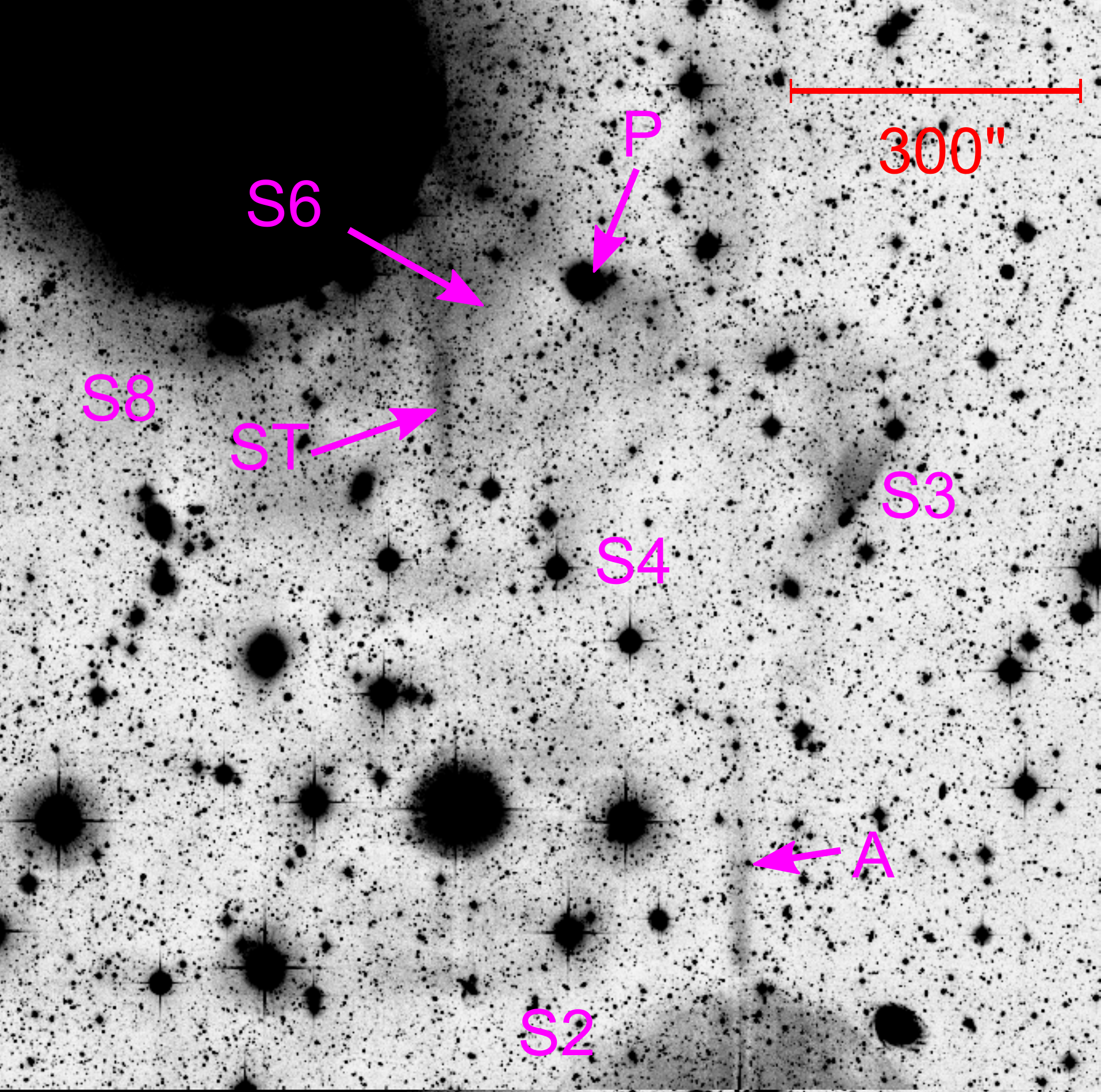}
\caption{MegaCam image. Minimum masking with a filter size of 44.4\arcsec. }
\label{fig:s6}
\end{figure} 

\begin{figure}
\includegraphics[width=\hsize]{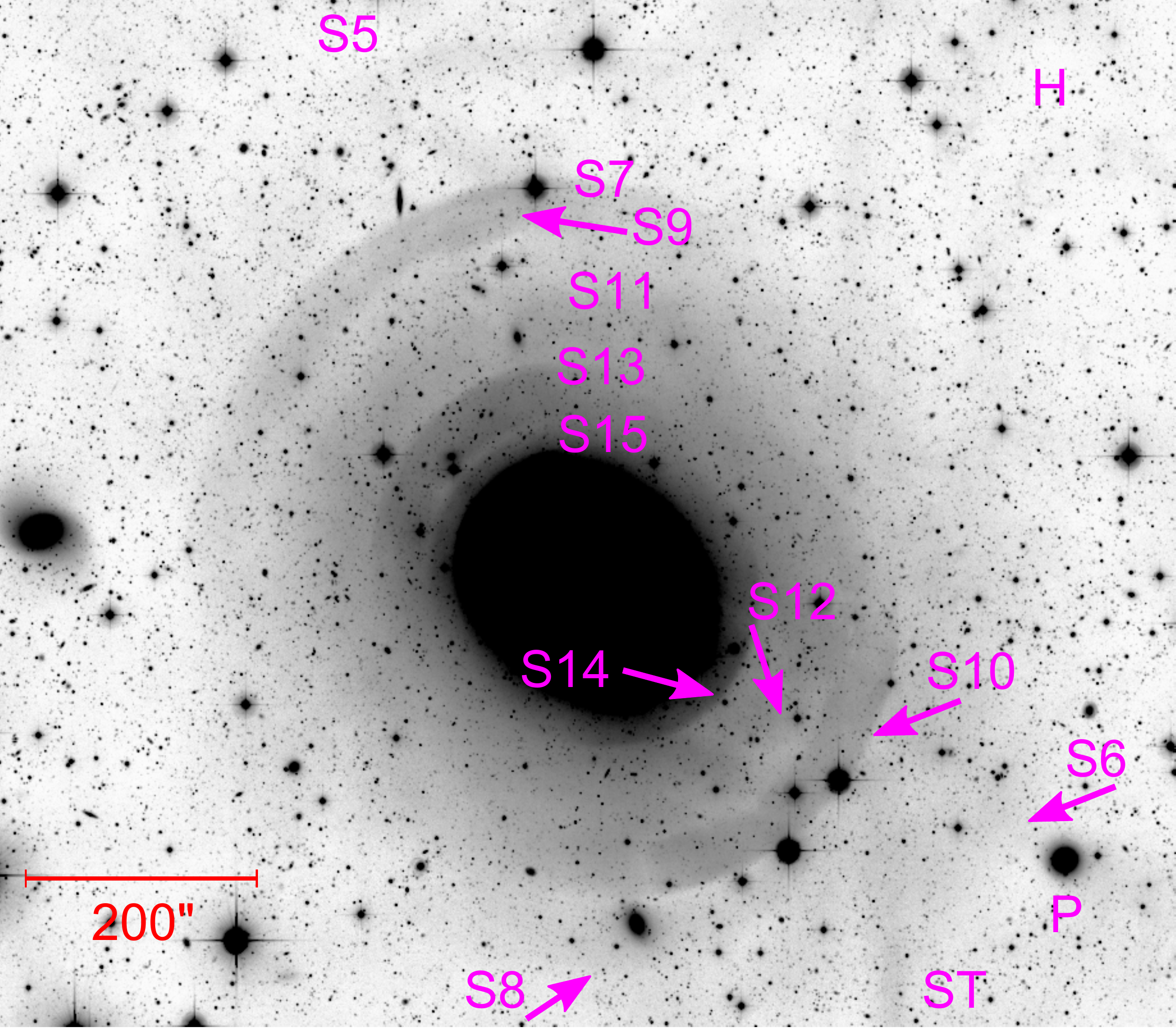}
\caption{MegaCam image. Minimum masking with a filter size of 44.4\arcsec.}
\label{fig:s9}
\end{figure} 

\begin{figure}[b!]\vspace{13.6ex}
\includegraphics[width=\hsize]{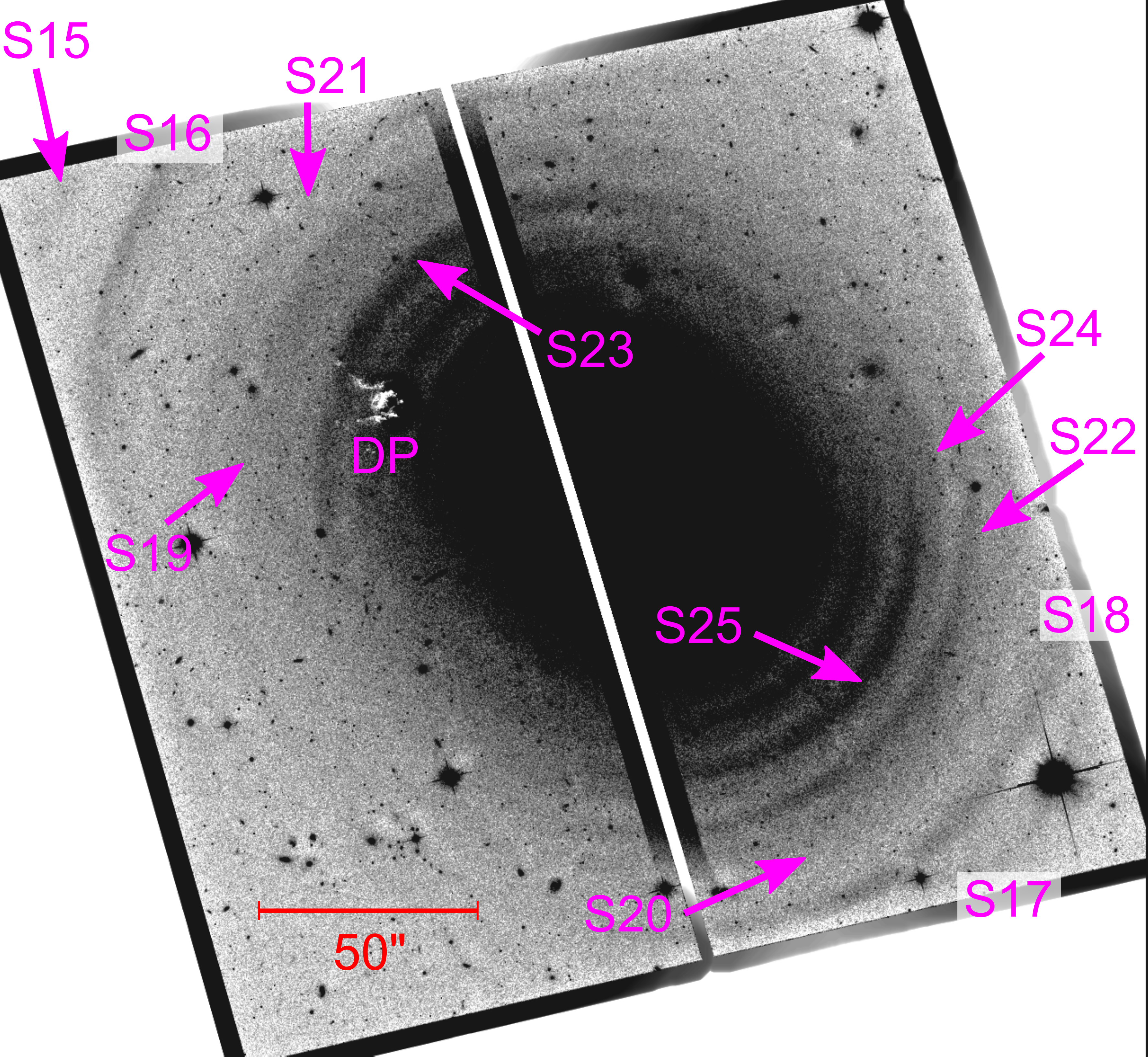}
\caption{HST image. Minimum masking with a filter size of 4\arcsec.}
\label{fig:s15}
\end{figure}


\begin{figure}
\includegraphics[width=\hsize]{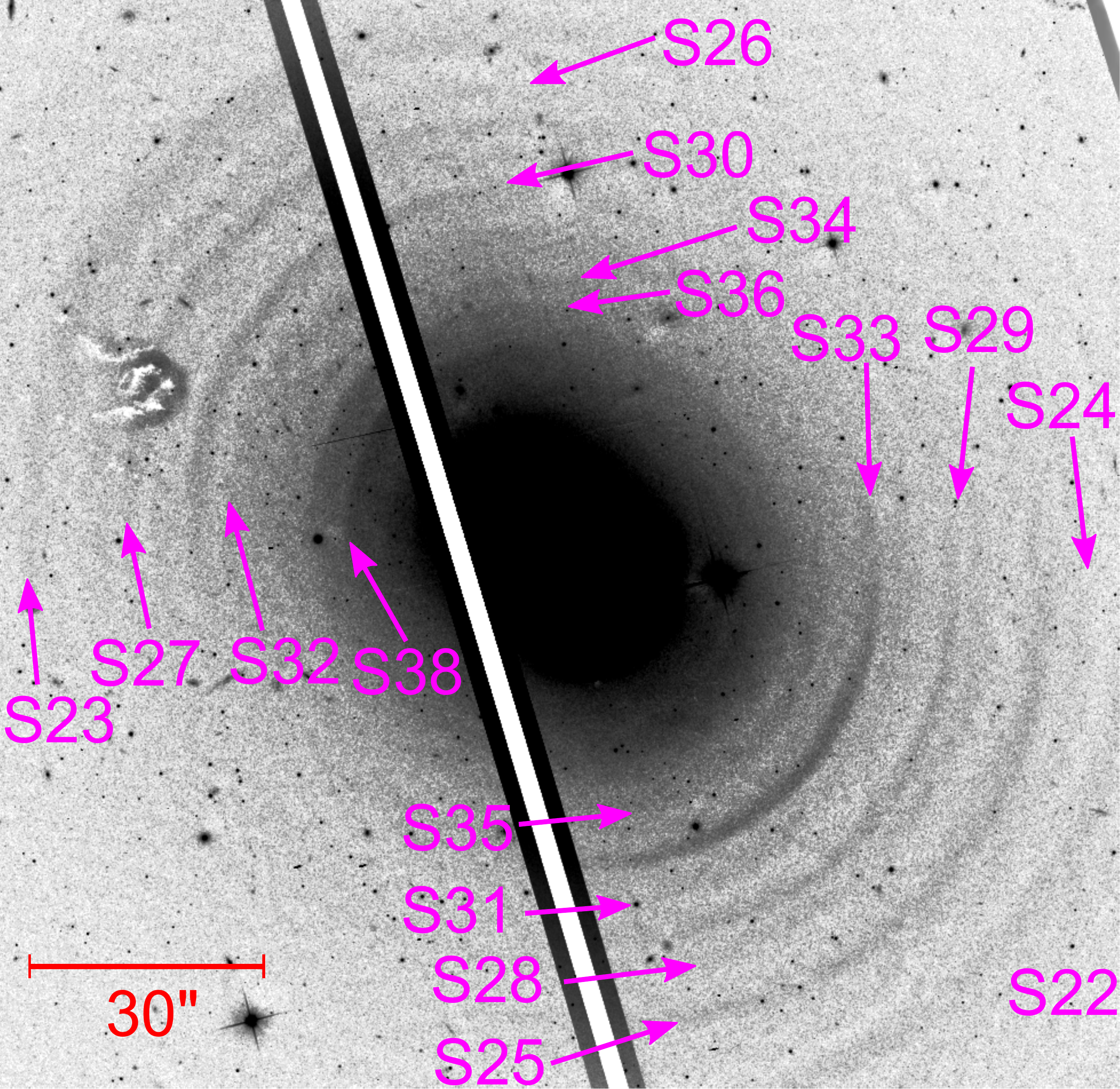}
\caption{HST image. Minimum masking with a filter size of 2\arcsec.}
\label{fig:s25}
\end{figure}

\begin{figure}\vspace{6ex}
\includegraphics[width=\hsize]{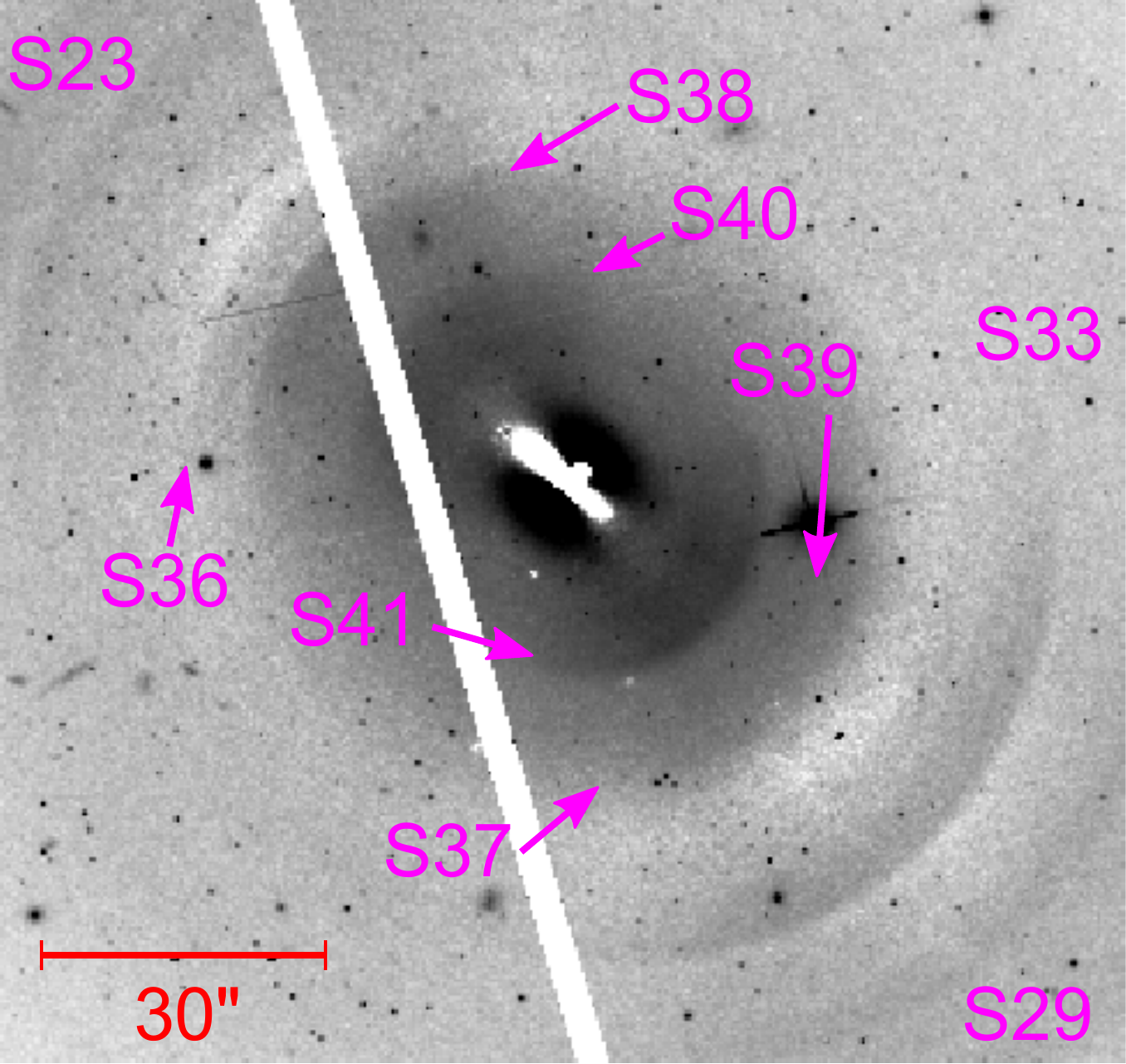}
\caption{HST image. Model subtraction.}
\label{fig:s39}
\end{figure} 

\begin{figure}
\includegraphics[width=\hsize]{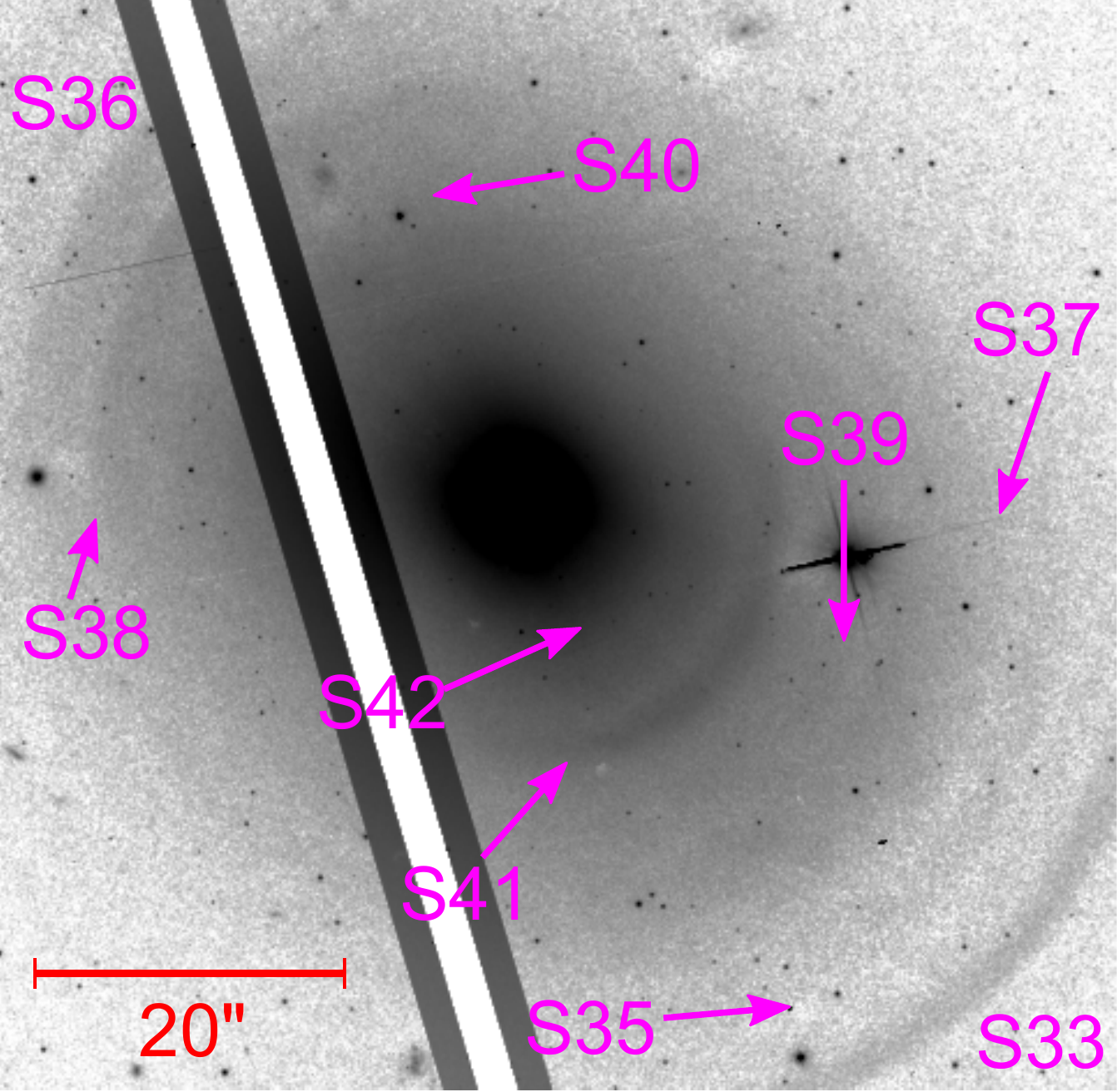}
\caption{HST image. Minimum masking with a filter size of 2\arcsec.}
\label{fig:s41}
\end{figure} 

\begin{figure}\vspace{5.8ex}
\includegraphics[width=\hsize]{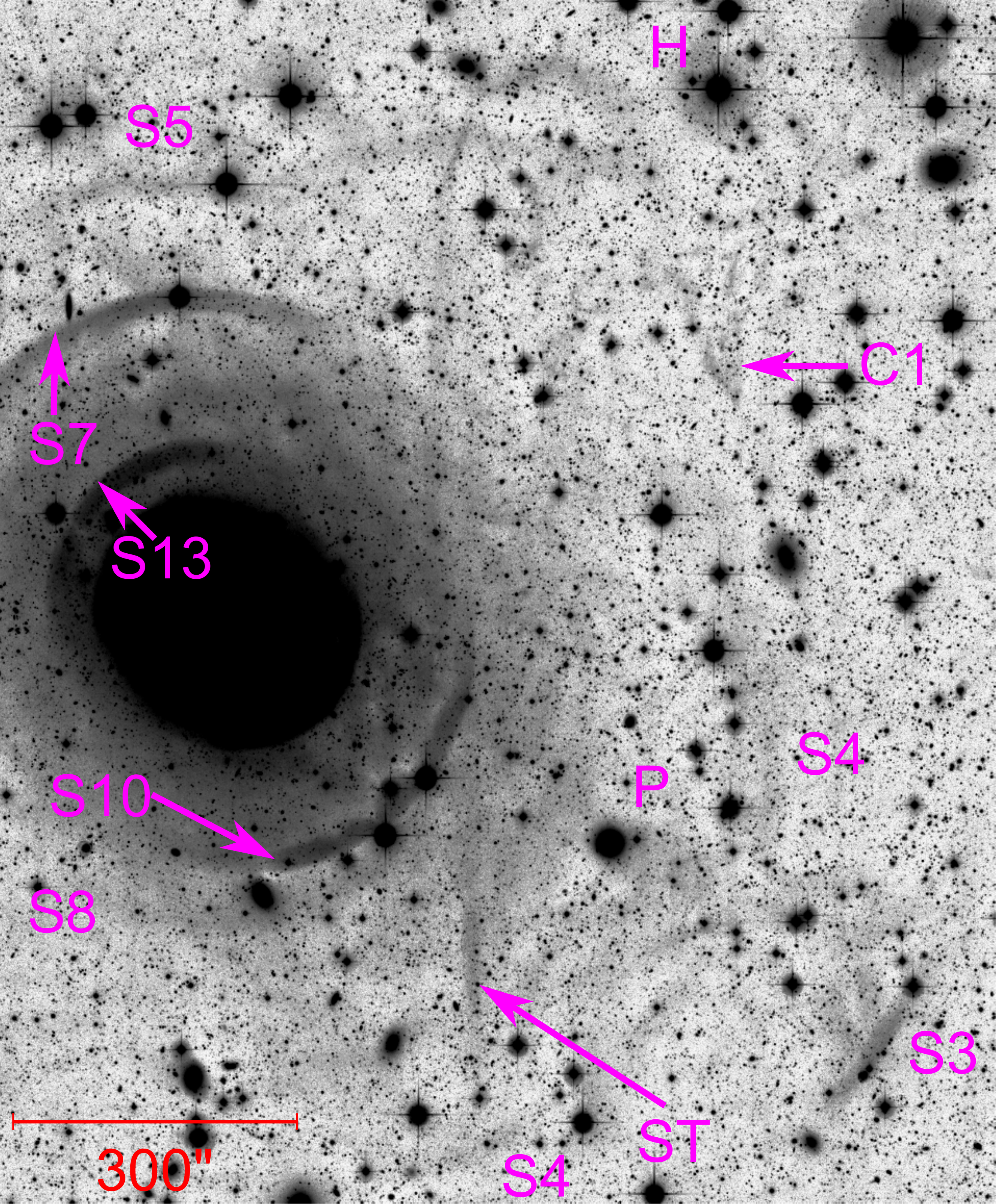}
\caption{The area of the hook. MegaCam image. Minimum masking with a filter size of 22.2\arcsec.}
\label{fig:hook}
\end{figure}

\begin{figure}
\includegraphics[width=\hsize]{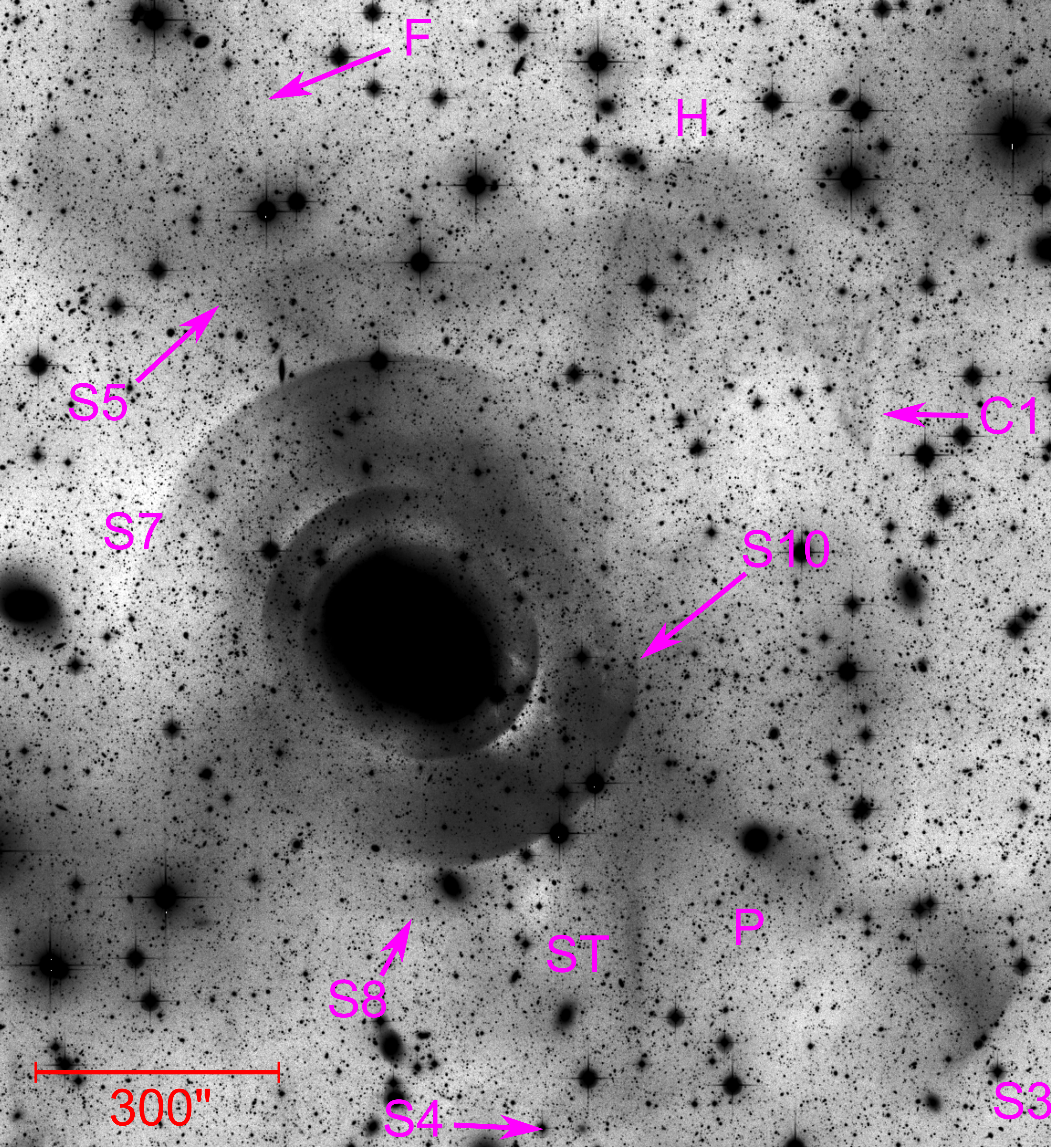}
\caption{Tidal features around NGC\,3923. MegaCam image. Model subtraction.}
\label{fig:tidal}
\end{figure}

\begin{figure}\vspace{6ex}
\includegraphics[width=\hsize]{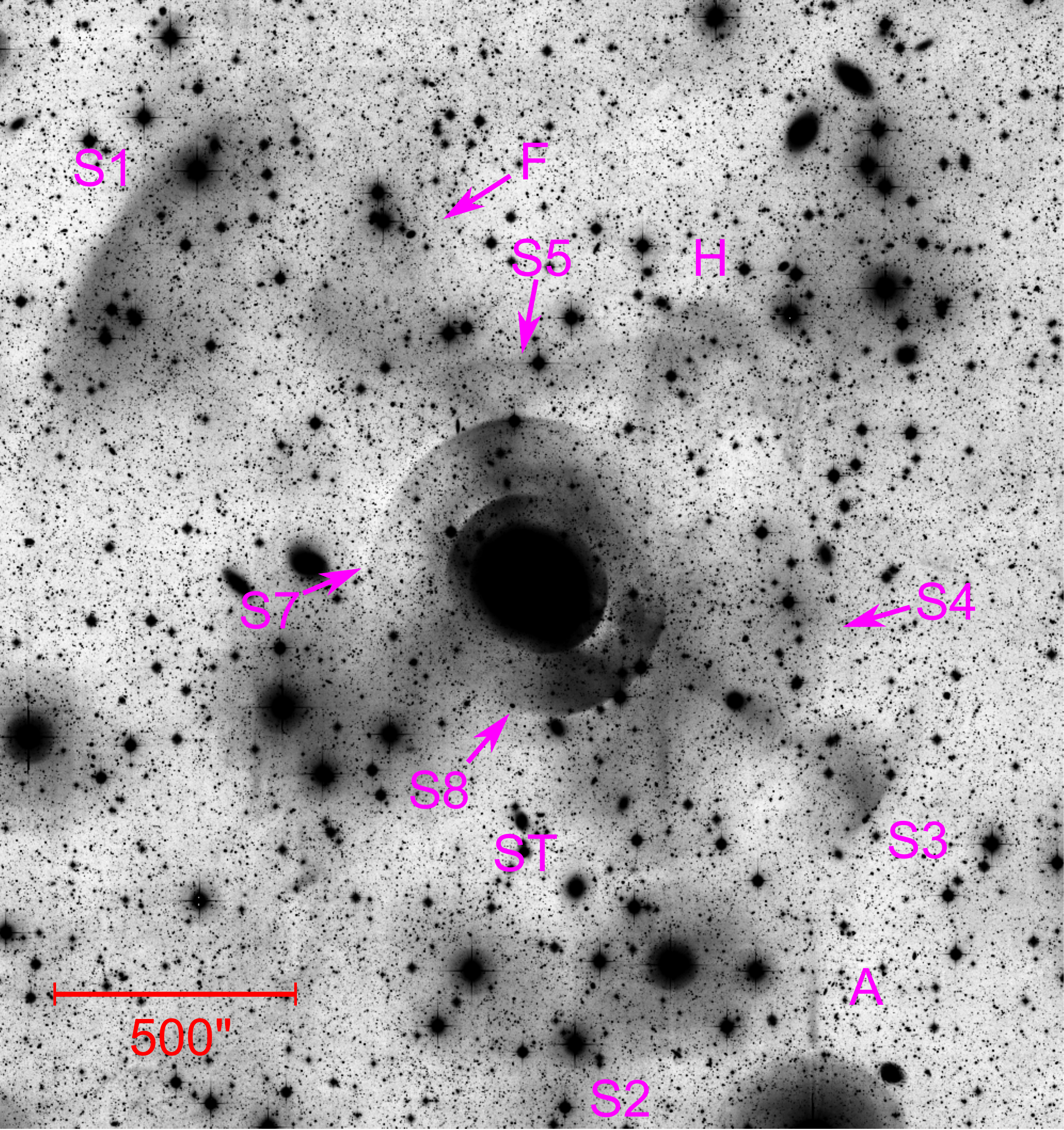}
\caption{Outskirts of NGC\,3923. MegaCam image. Model subtraction.}
\label{fig:outsk}
\end{figure}

\begin{figure}
\includegraphics[width=\hsize]{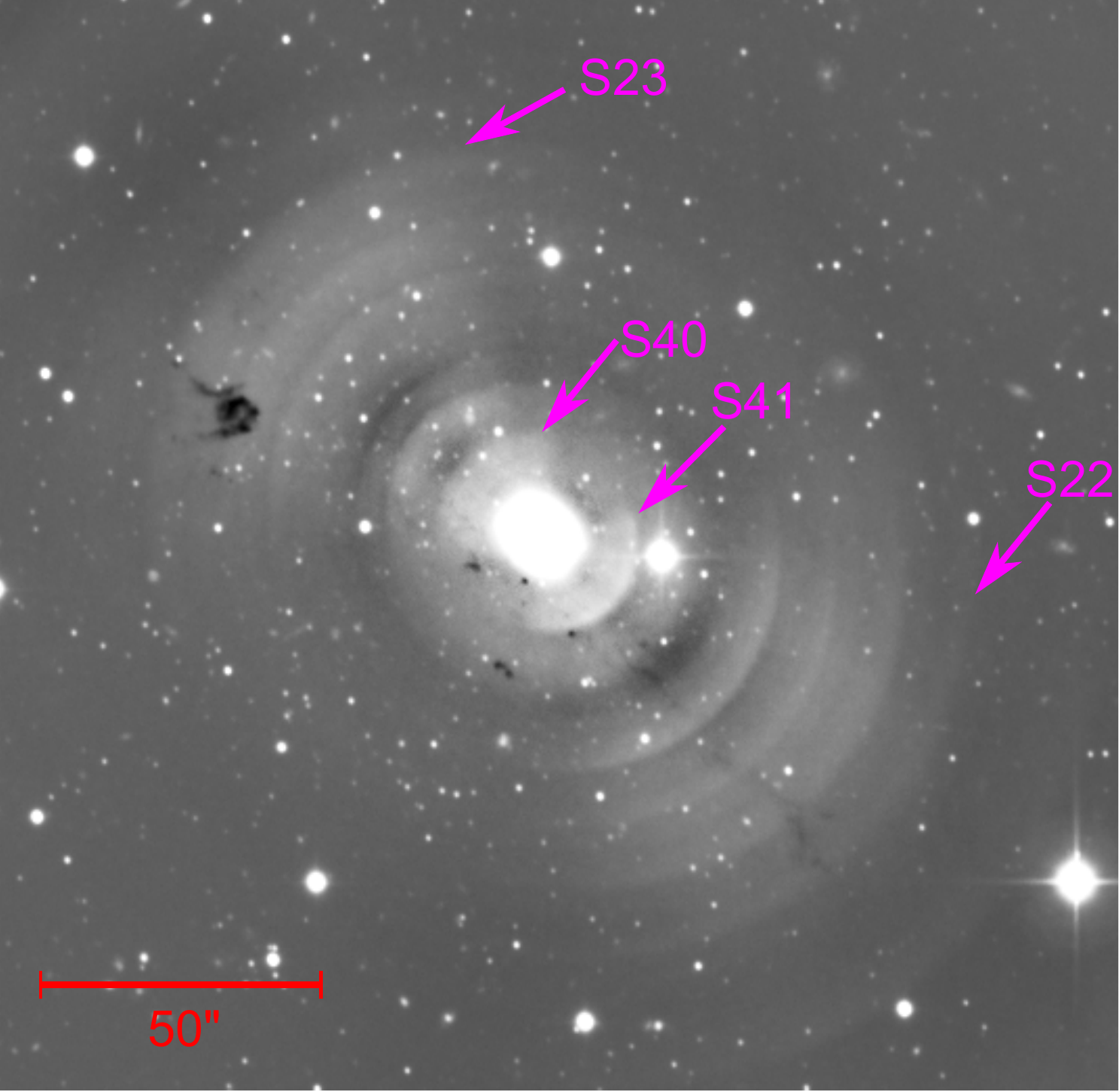}
\caption{Dust in the central parts of NGC\,3923. MegaCam image. Model subtraction.}
\label{fig:dust}
\end{figure}

\begin{figure}\vspace{12.2ex}
\includegraphics[width=\hsize]{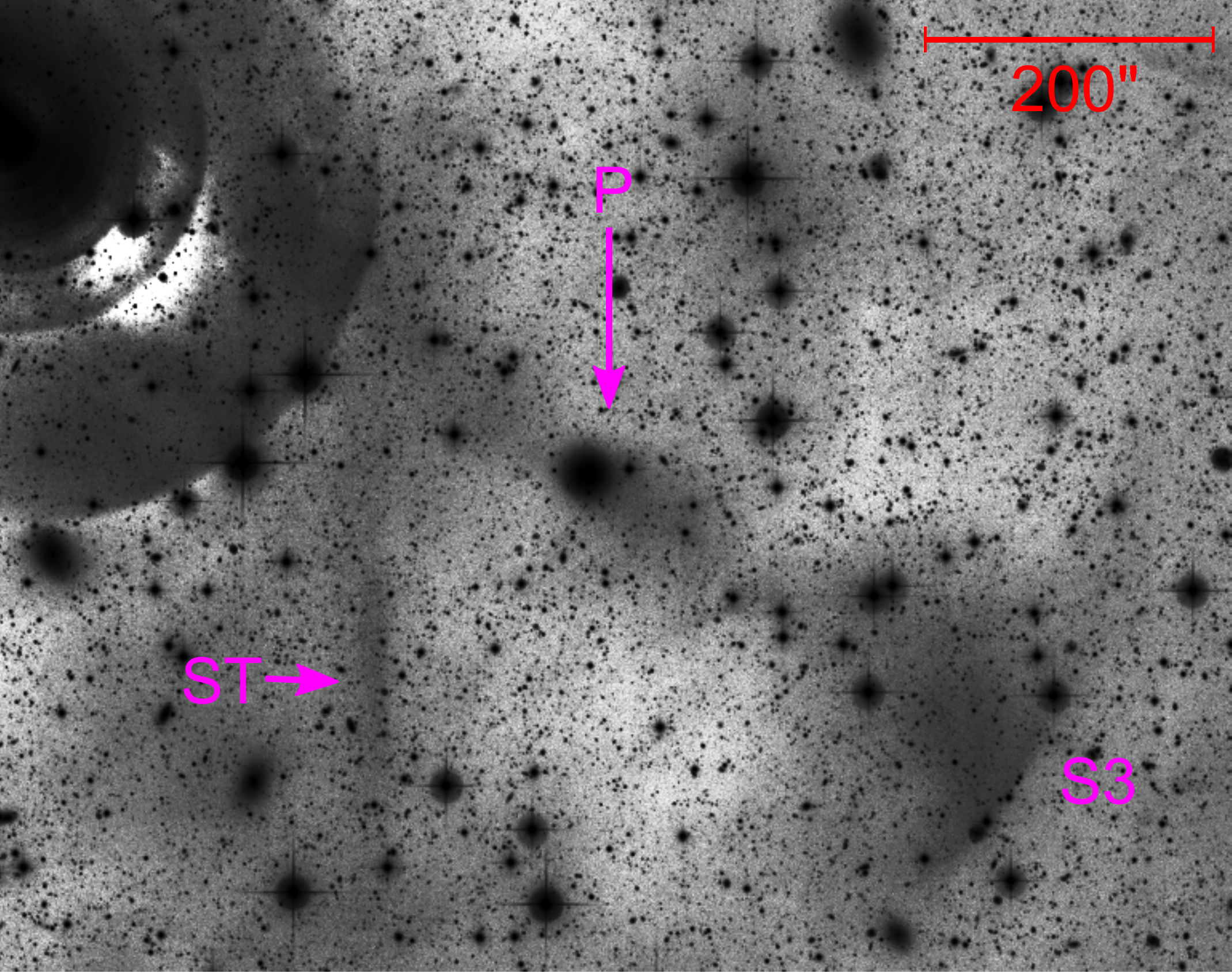}
\caption{Progenitor of some tidal features in NGC\,3923. MegaCam image. Model subtraction.}
\label{fig:progen}
\end{figure}

\clearpage

\section{Minimum masking}
\label{sec:minmask}
\begin{figure}[t!]
\includegraphics[width=0.97\hsize]{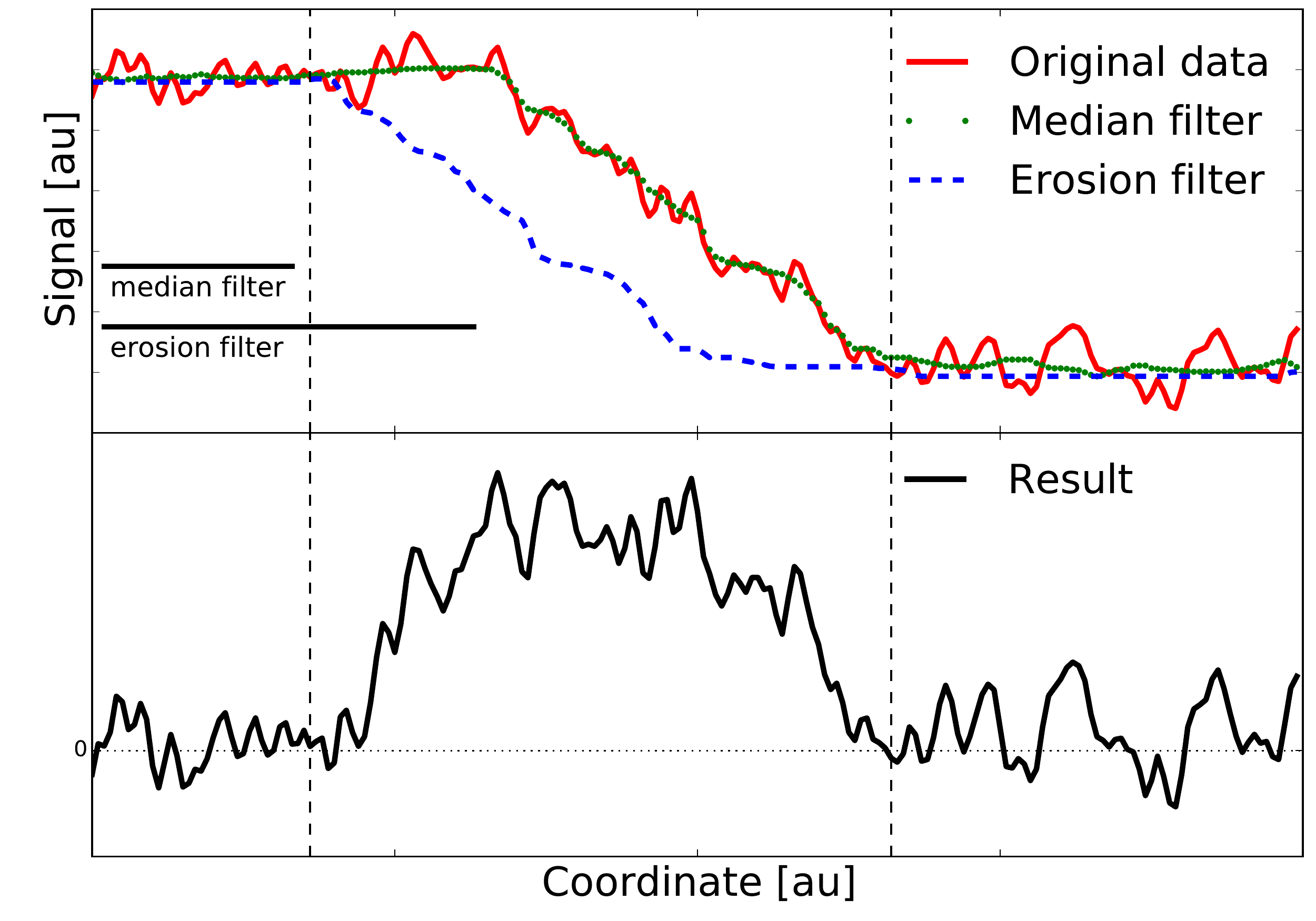}
\caption{Demonstration of the minimum masking on a~one-dimensional  example. Upper panel: The original data (red) are first smoothed by the median filter (green). Then the erosion (minimum) filter is applied (blue). Lower panel: The result is obtained as the difference between the original and the filtered data (the red and  blue curve, respectively). {The widths of the median and erosion filters are indicated in the figure. The unit \textit{au} means arbitrary unit.} The original (red curve) is a~step-like transition between two constant values with added noise. The resulting signal (black curve) oscillates around zero outside the area indicated by the vertical lines. The width of this area is determined by the width of the transition in the original signal and the chosen width of the filter. The original step transforms into a~bump. }
\label{fig:minmask}
\end{figure}

{Here we describe our minimum masking method for detecting the shells. The method is based on the operation known as erosion in image processing. The erosion operation is characterized by our choice of the {neighborhood} of every pixel which has a certain shape and size (rectangle, circle, line, etc.) and is the same for all the pixels in the image. At a given pixel, the result of erosion is the minimum over its neighborhood in the input image.  We used a circular neighborhood. To suppress the effects of noise, we smoothed the original image by the median filter before applying the erosion filter. For any pixel, the result of the median filter is the median over a neighborhood of the pixel.  Our median filter was square. The side of the square was the same as the radius of the circle we used for the minimum filter. This proved to be a simple strategy that worked  well. In summary, the minimum masking method means 1) using a median filter for the input image, and 2) applying erosion to the median filtered image;  the result is the difference between the input image and the eroded image.  The method is demonstrated in \fig{minmask} on a~one-dimensional example (the width of the erosion filter was twice the width of the median filter).}

{Minimum masking is closely related to the ``morphological gradient'' methods of edge detection in image processing. These methods are based on the difference between the original image and the eroded or dilated image. The dilatation operation is defined like the erosion, only the minimum is substituted by the maximum. Moreover, these methods implement a noise-smoothing step (e.g., \citealp{improc1} or \citealp{improc2}). Here the similarity with minimum masking ends. The image is then further processed to extract the edges.}

{We tested the ability of the minimum masking to detect shells in galaxy images. For this purpose, we made a restricted three-body simulation of a shell galaxy formation (see, e.g, \citealp{ebrova12} for details) by the most widely accepted shell formation mechanism -- the phase-wrapping minor merger model \citep{quinn83}. The two interacting galaxies were modeled as Plummer spheres colliding head-on. We stopped the simulation when several clearly defined shells developed. The particle positions were then projected into a plane containing the line of collision  to produce a surface density map (see \fig{sim}). This map was used subsequently for our tests. }

{In these tests, we were interested only in the rightmost shell in \fig{sim} to avoid the ambiguity in defining the   surface brightness of a shell because the other shells overlap each other in projection. The chosen shell was made of particles reaching their apocenters for the second time since the moment they were torn away from their mother galaxy. The radius of this reference shell was 135\,px (pixels).}

{We created two series of artificial images of shell galaxies. In the first series, the images were created as a sum of  the shell image from the simulation and of the image of the body of the galaxy modeled by a de~Vaucouleurs sphere. The half-light radius was chosen to be two times the radius of the reference shell. Then noise was added in the images. The new intensity of every pixel was a choice from the Poisson distribution whose mean value was proportional to the original intensity of the pixel. The test images in the second series were the same, but the shell image from the simulation was convolved by a Gaussian kernel with a standard deviation of 4\,px. This simulates the observed blurriness of shell edges (see the shell surface-brightness radial profiles in \citealp{sikkema07}). Each test image is characterized by 1) the ratio  of the brightest pixel of the reference shell and the corresponding underlying pixel of the body of the galaxy (Sh/B), and 2) the signal-to-noise ratio (S/N) at the same pixel.}

\begin{figure*}[b!]
\centering
  \begin{tabular}{@{}ccVVVV@{}}
    
    & & \multicolumn{4}{c}{\large{Signal-to-noise}}\\
    & & 100 & 50 & 10 & 5 \\  
      & $1/10$\! &
    \includegraphics[scale=1]{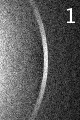}&
    \includegraphics[scale=1]{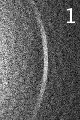} &
    \includegraphics[scale=1]{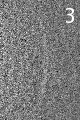} &
    \includegraphics[scale=1]{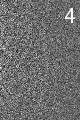} \\   
    \rotatebox[origin=c]{90}{\large{Shell-to-body}} & $1/50$ &   
    \includegraphics[scale=1]{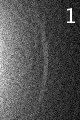} &
    \includegraphics[scale=1]{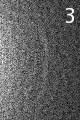} &
    \includegraphics[scale=1]{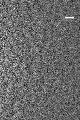} &
    \includegraphics[scale=1]{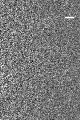} \\ 
    & $1/100$ &      
    \includegraphics[scale=1]{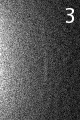} &
    \includegraphics[scale=1]{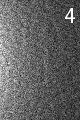} &
    \includegraphics[scale=1]{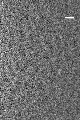} &
    \includegraphics[scale=1]{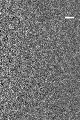} \\   
  \end{tabular}
  \caption{{Demonstration of the minimum masking method on an artificial image of a shell galaxy. The signal-to-noise ratio S/N varies in the horizontal direction. The ratio of the shell intensity to the galaxy body intensity Sh/B varies in the vertical direction.{ The number indicates our rating of prominence.}}}
  \label{fig:brsn1}
\end{figure*}

\begin{figure*}[b!]
\centering
  \begin{tabular}{@{}ccVVVV@{}}
    
    & & \multicolumn{4}{c}{\large{Signal-to-noise}}\\
    & & 100 & 50 & 10 & 5 \\  
      & $1/10$\! &
    \includegraphics[scale=1]{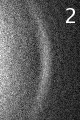} &
    \includegraphics[scale=1]{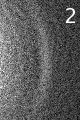} &
    \includegraphics[scale=1]{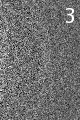} &
    \includegraphics[scale=1]{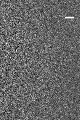} \\   
    \rotatebox[origin=c]{90}{\large{Shell-to-body}} & $1/50$ &   
    \includegraphics[scale=1]{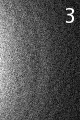} &
    \includegraphics[scale=1]{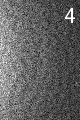} &
    \includegraphics[scale=1]{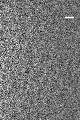} &
    \includegraphics[scale=1]{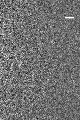} \\ 
    & $1/100$ &      
    \includegraphics[scale=1]{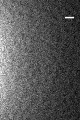} &
    \includegraphics[scale=1]{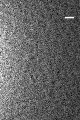} &
    \includegraphics[scale=1]{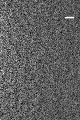} &
    \includegraphics[scale=1]{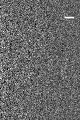} \\   
  \end{tabular}
  \caption{{Demonstration of the minimum masking method on an artificial image of a shell galaxy. The shell component was blurred to simulate the observed blurriness of shells. The signal-to-noise ratio S/N varies in the horizontal direction. The ratio of the shell intensity to the galaxy body intensity Sh/B varies in the vertical direction.{ The number indicates our rating of prominence.}}}
  \label{fig:brsn2}
\end{figure*}

{Finally, we applied minimum masking to the test images. The filter size for the first series was 5\,px and for the second, blurred, series 8\,px. These sizes proved to enhance the shell best irrespective of the Sh/B or S/N ratios. The results of the first series are shown in \fig{brsn1} and of the second series in \fig{brsn2}. The images show only the area of the shell of interest indicated by the red rectangle in \fig{sim}.}

\begin{figure*}[b!]
\centering
  \begin{tabular}{@{}cccc@{}}
     \multicolumn{4}{c}{\large{Galaxy-characteristic-radius-to-shell-radius}}\\
     1 & 2 & 5 & 10 \\  
    \includegraphics[scale=1]{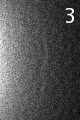} &
    \includegraphics[scale=1]{obr_test2/MinF_re2_SN100_rat100.png} &
    \includegraphics[scale=1]{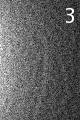} &
    \includegraphics[scale=1]{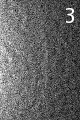} \\     
  \end{tabular}
  \caption{{Demonstration of the minimum masking method on an artificial image of a shell galaxy. The individual images differ by the ratio of the shell radius and the half-light radius of the galaxy body.{ The number indicates our rating of prominence.}}}
  \label{fig:shgr}
\end{figure*}

\begin{figure}[b!]
\includegraphics[width=\hsize]{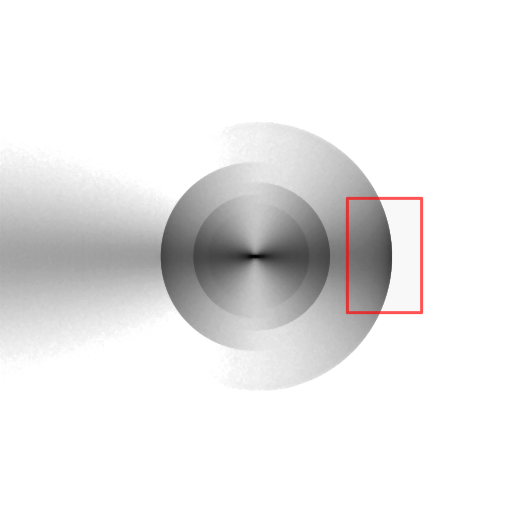}
\caption{{Simulation of a shell galaxy formation. Only the stars from the accreted galaxy are shown. The red rectangle marks the region displayed in Figs.~\ref{fig:brsn1}--\ref{fig:shgr}.}}
\label{fig:sim}
\end{figure}

{In the two basic series of test images, the ratio of the reference shell radius and the characteristic radius of the de~Vaucouleurs sphere was fixed. In \fig{shgr}, we also varied the characteristic radius of the body of the galaxy. We can see that the ratio of the shell and the galaxy radius has only a mild influence on the shell detectability by the minimum masking method.}

\begin{figure*}
\begin{tabular}{ccc}
\subfloat[Minimum masking -- filter size 4\arcsec]{\includegraphics[width = 2in]{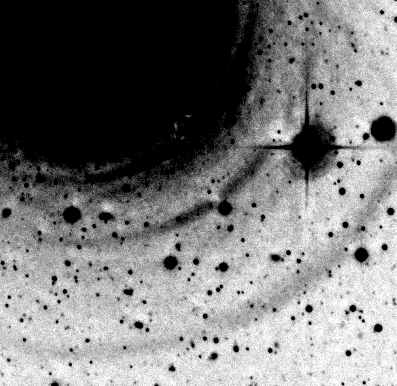}} &
\subfloat[Minimum masking -- filter size 8\arcsec]{\includegraphics[width = 2in]{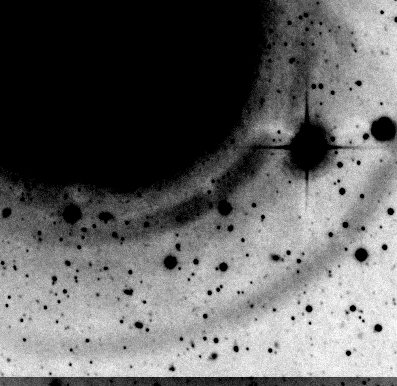}} &
\subfloat[Minimum masking -- filter size 16\arcsec]{\includegraphics[width = 2in]{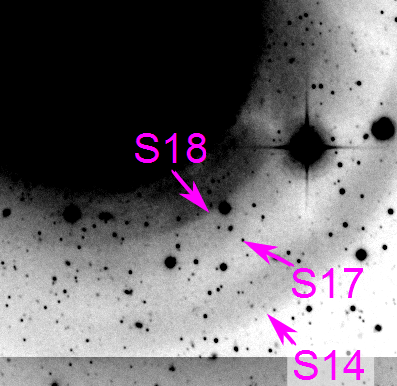}} \\
\subfloat[Linear scaling]{\includegraphics[width = 2in]{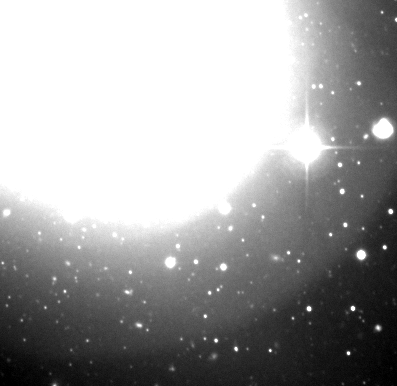}} &
\subfloat[Model subtraction]{\includegraphics[width = 2in]{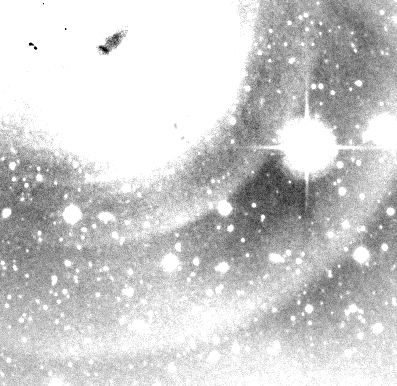}} &
\subfloat[Unsharp masking -- filter size 4\arcsec]{\includegraphics[width = 2in]{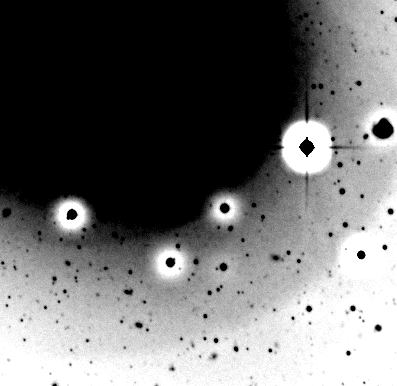}} 
\end{tabular}
\caption{{Detecting  shell S17 by various methods. MegaCam image.}}
\label{fig:shdet}
\end{figure*}

{We also compared various methods of shell detection on our MegaCam image of NGC\,3923 to enhance the shell S17 (see \fig{shdet}). The upper row of tiles shows the image processed by the minimum masking method with various filter sizes. The shell is clearly detected. Comparing the images processed by the various filter sizes, we can easily recognize  that the shell S17 is not an artifact. Tile (d) is processed by a simple linear scaling of the original image. Shell S17 is still clearly detectable here. In  tile (e), a smooth two-component S{\' e}rsic model of the galaxy body was subtracted. The large-scale variations make the detection of  shell S17 difficult. These variations  are caused by the deviations of the real galaxy profile from the model. Tile (f) is the image processed by unsharp masking. The mask had a Gaussian kernel with a standard deviation of 4\arcsec. This size proved to be the best, but even so the shell in question can barely be seen. }

\newcolumntype{V}{>{\centering\arraybackslash} m{.14\linewidth} }

\end{appendix}

\end{document}